\documentclass[twocolumn,prl,showpacs,nofootinbib]{revtex4}

\usepackage{graphicx}
\usepackage{dcolumn}
\usepackage{bm}
\def\veck{\bm{k}}
\def\vecq{\bm{q}}

\def\hatk{\bm{\hat k}}

\def\VEV#1{\left\langle #1 \right\rangle}
\begin{document}

\title{Nonlinear Evolution of Anisotropic Cosmological Power}

\author{Shin'ichiro Ando and Marc Kamionkowski}
\affiliation{California Institute of Technology, Mail Code 130-33,
Pasadena, CA 91125}

\date{November 5, 2007; revised January 8, 2008; accepted January 10, 2008}

\begin{abstract}
There has been growing interest in the possibility of testing
more precisely the assumption of statistical isotropy of
primordial density perturbations.  If it is to be tested with galaxy
surveys at distance scales $\alt 10~{\rm Mpc}$, then
nonlinear evolution of anisotropic power must be understood.  To
this end, we calculate the angular dependence of the power
spectrum to third order in perturbation theory for a primordial
power spectrum with a quadrupole dependence on the wavevector
direction.  Our results suggest that primordial power
anisotropies will be suppressed by $\alt 7$\% in the quasilinear
regime.  We also show that the skewness in the statistically
anisotropic theory differs by no more than 1\% from that in the
isotropic theory.
\end{abstract}


\maketitle

It is commonly assumed that primordial density perturbations are
statistically isotropic, and statistical isotropy is a
prediction of most structure-formation theories.  The notion of
statistical isotropy can be quantified, though, by considering
models in which it is broken, and a growing
literature has discussed physical models that produce primordial
perturbations that are not statistically isotropic
\cite{preferreddirections}.  

The manifestations of departures from statistical isotropy can
take on many forms, but one simple example \cite{Ackerman:2007nb}
predicts a primordial power spectrum $P(\veck)$ with a quadrupole
dependence on the direction $\hatk$ of the wavevector $\veck$,
\begin{equation}
 P(\veck) = A(k)
  \left[1 + g_\ast \mathcal P_2(\mu_k) \right],
\label{eq:anisotropic power spectrum}
\end{equation}
where $\mathcal P_2(x)=(3x^2-1)/2$ is the second Legendre
polynomial, $\mu_k = \hatk \cdot \bm{\hat e}$, and $\bm{\hat e}$ is a
preferred direction.\footnote{Note that our $g_\ast$ is $3/2$ times that
in Ref.~\cite{Ackerman:2007nb}.}
Reference~\cite{Pullen:2007} constructed a CMB minimum-variance
estimator for the power-anisotropy amplitude $g_\ast$ and showed that
the Planck satellite will be sensitive to a value of $g_\ast$ as small
as $\sim$0.02, a number which can be estimated roughly by
$\sim$10\,$N_{\rm pix}^{-1/2}$, where $N_{\rm pix}$ is the number
of statistically independent pixels on the sky detected by
Planck.

Similar arguments suggest that the sensitivity of a galaxy
survey, like the Sloan Digital Sky Survey \cite{sdss} or
Two-Degree Field \cite{2dF}, should have a
comparable sensitivity.  One issue that will arise, though, in
testing statistical isotropy of primordial perturbations is the
quasilinear evolution of the power spectrum.  The root-mean-square
density-perturbation amplitude becomes of order unity at distance scales
$\sim$10\,$h^{-1}\,{\rm Mpc}$, and so quasilinear evolution of
the density field must be taken into account if the mass
distribution measured on these scales in the Universe today is
to be used to infer the primordial mass distribution.

In this paper, we study the nonlinear evolution of density perturbations
to see how nonlinearity affects statistical isotropy.  Does nonlinear
evolution amplify departures from statistical isotropy? Or possibly
suppress them?  To take the first steps to address this question, we
calculate the power spectrum, to third order in perturbation
theory, under the assumption that primordial perturbations have
the form given in Eq.~(\ref{eq:anisotropic power spectrum}).  
The bottom line is that quasilinear evolution {\it suppresses}
departures from statistical isotropy, but by only $\alt 7$\%
compared with the linear theory.  Thus, galaxy surveys
in the linear/quasilinear regime will still be useful diagnostics for
departures from statistical isotropy.  We also calculate skewness,
finding that the quadrupole power anisotropy changes it by no more than
1\%.

To proceed, we use third-order perturbation theory to determine
whether a primordial power anisotropy is amplified or suppressed
in the quasilinear regime.  
The power spectrum $P(\veck, z)$ is defined by the ensemble average of
the two-point correlation of the Fourier transform $\hat \delta
(\veck, z)$ of the density perturbation through
\begin{equation}
      \langle \hat\delta (\veck_1, z) \hat\delta (\veck_2, z) \rangle
        = (2 \pi)^3 \delta_D (\veck_1 + \veck_2) P(\veck_1,z),
  \label{eq:power spectrum}
\end{equation}
where $\delta_D$ is the Dirac delta function.
The density perturbations can be expanded in terms of the
linear-theory density-perturbation amplitude, which has a redshift
dependence proportional to the linear growth factor $D(z)$,
\begin{equation}
 \hat \delta (\veck, z) = \sum_{n=1}^{\infty}
  \delta_n (\veck) D^n(z).
  \label{eq:density expansion}
\end{equation}
In the linear regime, each Fourier mode grows at the same rate,
and so the linear-theory power spectrum $P_{\rm lin}(\veck)$,
corresponding to the linear density perturbation $\delta_1
(\veck)$, evolves in such a way that the quadrupole dependence
of the primordial power spectrum [Eq.~(\ref{eq:anisotropic power
spectrum})] is preserved.  However, the nonlinear power spectrum
will have different anisotropy structure, and we shall
investigate it following prescription developed in
Refs.~\cite{Goroff:1986,Makino:1992,Jain:1994}.

Reference~\cite{Goroff:1986} showed that solution of
the nonrelativistic fluid equations (i.e., continuity, Euler,
and Poisson equations) allows one to write the $n$th-order
density-perturbation amplitude $\delta_n$ in terms of the linear
perturbation $\delta_1$ through
\begin{eqnarray}
 \delta_n(\veck) & = & \int \frac{d\vecq_1}{(2\pi)^3} \cdots
  \int \frac{d\vecq_n}{(2\pi)^3}
  \delta_D(\vecq_1 + \cdots + \vecq_n - \veck)
  \nonumber\\&&{}\times
  (2\pi)^3 F_n (\vecq_1,\cdots,\vecq_n)
  \delta_1(\vecq_1)\cdots \delta_1(\vecq_n).
  \label{eq:delta_n}
\end{eqnarray}
For any $n$, the function $F_n$ can be obtained by recursive relations
given in Ref.~\cite{Goroff:1986}.  In particular, the expressions for
$F_2$ and $F_3$ that are directly relevant for our purpose are
explicitly given by Eqs.~(A2) and (A3) of
Ref.~\cite{Goroff:1986}.\footnote{Strictly speaking, $F_2$ in
Ref.~\protect\cite{Goroff:1986} is for an Einstein-de Sitter
Universe, but Ref.~\cite{Kamionkowski:1998fv} shows that its form
in a $\Lambda$CDM model with $\Omega_m\simeq 0.3$ differs by
less than 1\%; the same should be true for $F_3$.}
As all odd moments of $\delta_1 (\veck)$ vanish, the power spectrum to
third order is given as
\begin{eqnarray}
      \lefteqn{
      (2 \pi)^3 \delta_D (\veck_1 + \veck_2) P^{(2)}(\veck_1, z)}
       \nonumber\\ &=&
        D^2(z) \langle \delta_1(\veck_1) \delta_1(\veck_2) \rangle
         + D^4(z) \left[ \langle \delta_2(\veck_1) \delta_2(\veck_2) \rangle
          \right.\nonumber\\&&{}\left.
           + \langle \delta_1(\veck_1) \delta_3(\veck_2) \rangle
            + \langle \delta_3(\veck_1) \delta_1(\veck_2) \rangle \right].
   \label{eq:third-order power spectrum}
\end{eqnarray}
The second term, which evolves as $D^4(z)$, is the next-to-leading term
evaluated to the third order of density perturbation $\delta_3$, while the
first term is the linear part.  We further define a quantity $P_{mn}$ as
\begin{equation}
    \langle \delta_m(\veck_1) \delta_n(\veck_2) \rangle
     = (2 \pi)^3 \delta_D(\veck_1 + \veck_2) P_{mn}(\veck_1),
\end{equation}
and with this definition, Eq.~(\ref{eq:third-order power
spectrum}) is rewritten as
\begin{equation}
 P^{(2)}(\veck,z)  =  D^2(z) P_{11}(\veck)
  + D^4(z) [P_{22}(\veck) + 2 P_{13}(\veck)].
  \label{eq:ps}
\end{equation}

We then need to write $P_{22}$ and $P_{13}$ in terms of the linear
spectrum $P_{\rm lin} = P_{11}$.
This procedure is straightforward, and the results are very similar to
those given in Ref.~\cite{Jain:1994}:
\begin{eqnarray}
 P_{22}(\veck) & = & 2 \int \frac{d\vecq}{(2\pi)^3}
  \left[F_2^{(s)}(\vecq, \veck - \vecq)\right]^2
  \nonumber\\&&{}\times
  P_{\rm lin}(\vecq) P_{\rm lin}(\veck - \vecq),
  \label{eq:P_22}
 \\
 P_{13}(\veck) & = & 3 P_{\rm lin}(\veck)
  \int \frac{d\vecq}{(2\pi)^3} F_3^{(s)}(\vecq, -\vecq, \veck)
  P_{\rm lin}(\vecq),
  \label{eq:P_13}
\end{eqnarray}
where the symmetrized function $F_n^{(s)}$ is obtained by summing $n!$
permutations of $F_n$ over its $n$ arguments and dividing by $n!$.
The only difference from the expressions in Ref.~\cite{Jain:1994} is
that the linear power spectrum now depends on both magnitude $k$
and direction $\hatk$ of wavevector $\veck$.

In order to illuminate the anisotropy structure, we expand
Eqs.~(\ref{eq:P_22}) and (\ref{eq:P_13}) with Legendre
polynomials.  The most general power spectrum today ($D = 1$)
can be expanded in Legendre polynomials ${\mathcal P}_n(x)$ as
\begin{equation}
  P(\veck)  =  P_{\rm lin}(k,\mu_k) + \sum_{m,n=0}^{\infty}
  g_\ast^m \mathcal P_n (\mu_k) B_{mn}(k),
\label{eq:completepowerspectrum}
\end{equation}
where the sum on $n$ is restricted to even positive integers.
To the next-to-leading order, the expansion can be written,
\begin{eqnarray}
P^{(2)}(\veck)  &=& A_{\rm lin}(k) + B_{00}(k) + g_\ast^2 B_{20}(k)
    \nonumber\\&&{}
  +\left\{ g_\ast [ A_{\rm lin}(k)
    + B_{12}(k) ]
    +g_\ast^2 B_{22}(k) \right\} \mathcal P_2(k)
    \nonumber\\&&{}
  +g_\ast^2 B_{24}(k) \mathcal P_4(k),
 \label{eq:third-order power spectrum 2}
\end{eqnarray}
where $A_{\rm lin}$, defined by Eq.~(\ref{eq:anisotropic power
spectrum}), is the isotropic linear power spectrum.
Expressions for the expansion coefficients $B_{mn}(k)$ are given
at the end of the paper.  We then rewrite
Eq.~(\ref{eq:third-order power spectrum 2}) as
\begin{eqnarray}
 P^{(2)}(\veck) & = & A^{(2)}(k) \left[1 + g_2^{(2)}(k)
     \mathcal P_2(\mu_k) 
    \right.\nonumber\\&&{}\left.
    + g_4^{(2)}(k) \mathcal P_4(\mu_k) \right],
 \label{eq:third-order power spectrum 3}
 \\
 g_2^{(2)}(k) &=& g_\ast c_1(k) + g_\ast^2 c_2(k),
 \label{eq:g_2}
 \\
 g_4^{(2)}(k) &=& g_\ast^2 c_3(k),
 \label{eq:g_4}
\end{eqnarray}
with the following definitions of $A^{(2)}$, $c_1$, $c_2$, and $c_3$:
\begin{eqnarray}
  A^{(2)}(k) & = &  A_{\rm lin}(k) + B_{00}(k) + g_\ast^2 B_{20}(k),
  \\
  A^{(2)}(k)c_i(k) &=& \left\{
  \begin{array}{cc}
    A_{\rm lin}(k) + B_{12}(k) & [i=1] \\
    B_{22}(k) & [i=2] \\
    B_{24}(k) & [i=3] \\
  \end{array}\right..
  \label{eq:c}
\end{eqnarray}
We note that the primordial quadrupole anisotropy also affects the
isotropic part of the next-to-leading order power spectrum, $A^{(2)}(k)$.
While this is interesting, that correction is expected to be very small,
being suppressed by $g_\ast^2$.
Thus, hereafter we neglect this term; keeping it only gives correction
of the order of $g_\ast^3$ and $g_\ast^4$ to $g_2^{(2)}$ and
$g_4^{(2)}$, respectively.

\begin{figure}
\begin{center}
\includegraphics[width=8.5cm]{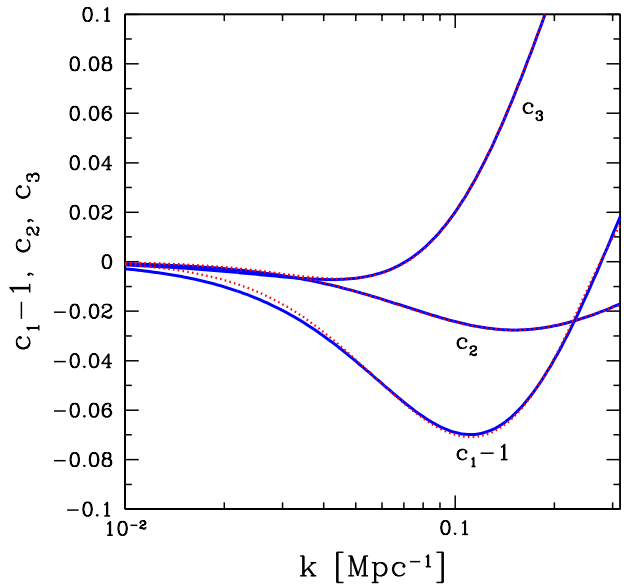}
\caption{Coefficients of anisotropic terms in the third-order power
  spectrum, $c_1(k)-1$, $c_2(k)$, and $c_3(k)$.  The definitions are
  given in Eqs.~(\ref{eq:third-order power spectrum
  3})--(\ref{eq:g_4}).  Solid curves are the result of numerical
  integration, while dotted curves are the fitting functions
  (\ref{eqn:ansatz 1})--(\ref{eqn:ansatz 3}).}
\label{fig:coeff}
\end{center}
\end{figure}

For the linear power spectrum $A_{\rm lin}(k)$, we use the
fitting formula for the transfer function given in
Ref.~\cite{Eisenstein:1999} with current values for the relevant
cosmological parameters \cite{Spergel:2006}.

If $g_\ast \ll 1$, we may neglect all the terms proportional to
$g_\ast^2$, and thus $c_1(k)$ is the only quantity relevant for
anisotropy structure.
As linear theory simply gives $c_1(k) = 1$, the quantity $c_1(k)-1$
represents enhancement of the quadrupole anisotropy due to quasilinear
evolution.
In Fig.~\ref{fig:coeff}, we plot $c_1(k) - 1$.
This figure shows that at large scales $k \alt 10^{-2}$ Mpc$^{-1}$,
the nonlinear effect is subdominant and thus anisotropic structure is
the same as that for the linear theory: $c_1(k) \approx 1$.
The quadrupole anisotropy then decreases at smaller scales and it
reaches minimum ($c_1 - 1 \simeq -0.07$) in the quasilinear regime, $k \sim
0.1$ Mpc$^{-1}$.

Third-order perturbation theory becomes less accurate for even
higher wavenumbers, $k \agt 0.1$ Mpc$^{-1}$, at $z = 0$.
However, it remains very accurate at such (comoving) scales in
the high-redshift Universe, $z \ge 1$~\cite{Jeong:2006xd}.
For general redshifts $z$, the anisotropic power spectrum is given by
Eqs.~(\ref{eq:third-order power spectrum 3})--(\ref{eq:c}) with
replacements $A_{\rm lin} \to D^2(z) A_{\rm lin}$ and $B_{mn} \to
D^4(z) B_{mn}$.
The enhancement of anisotropy is then given by $c_1(k,z) - 1$, which
is well approximated by $[c_1(k) - 1] D^2(z)$ in the quasilinear regime
where $A_{\rm lin} > |B_{mn}|$.
It is $-0.03$ at $k = 0.1$ Mpc$^{-1}$ and $z = 1$; the suppression is
weaker.
One can also see amplified anisotropy for $k \agt 0.3$ Mpc$^{-1}$ at
redshifts when the third-order approach is still valid at such scales.

The corrections on the order of $g_\ast^2$ are represented by $c_2(k)$
and $c_3(k)$, both of which are also shown in Fig.~\ref{fig:coeff}.
Besides it is suppressed by additional $g_\ast$, $c_2(k)$ is
intrinsically smaller than $c_1 - 1$, thus giving only a minor
correction.
While possessing different anisotropy structure, the higher multipole
term proportional to $c_3(k)$ would also be small in the quasilinear
regime.

The results of our computations can be approximated by
\begin{eqnarray}
     c_1(k) &=& 1-0.463 \Delta^2(k) + 0.886 \Delta^4(k)
     -0.407 \Delta^6(k),
     \label{eqn:ansatz 1} \\
     c_2(k) &=& -0.133 \Delta^2(k) + 0.195 \Delta^4(k) - 0.079
     \Delta^6(k),
     \label{eqn:ansatz 2}\\
     c_3(k) &=& -0.163 \Delta^2(k) + 1.062 \Delta^4(k) -1.082
     \Delta^6(k)
     \nonumber\\&&{}
     + 0.373 \Delta^8(k),
     \label{eqn:ansatz 3}
\end{eqnarray}
in terms of the linear-theory density-perturbation amplitude at
wavenumber $k$, $\Delta^2(k)=(k^3/2\pi^2) A_{\rm lin}(k)$.
These fitting functions are plotted as dotted curves in
Fig.~\ref{fig:coeff}.  They were obtained explicitly for the
current best-fit cosmological parameters \cite{Spergel:2006},
but when written in terms of $\Delta^2(k)$, should also be
accurate for other cosmological parameters.  Strictly speaking,
the validity of third-order perturbation theory breaks down
when the terms higher order in $\Delta^2(k)$ become important;
in this case, one will need to go to fourth-order or even
higher-order corrections to obtain the evolved power spectrum.

We now compute the skewness $S_3 \equiv \VEV{\delta^3} /
\VEV{\delta^2}^2$ from the anisotropic primordial power spectrum, to
second order in perturbation theory.  The standard result, for an
Einstein-de Sitter Universe, is $S_3=34/7$ \cite{pee80,Fry:1983cj}, and
the result for a $\Lambda$CDM Universe differs by less than 1\%.
The three-point correlation function at zero lag
is a Fourier transform of $\langle \hat \delta
(\veck_1, z) \hat \delta (\veck_2, z) \hat \delta (\veck_3, z)
\rangle$, and thus the
leading contribution comes from $D^4(z) \langle \delta_1(\veck_1)
\delta_1 (\veck_2) \delta_2 (\veck_3) \rangle$, etc.
Rewriting $\delta_2$ in terms of the linear fluctuation $\delta_1$ and the
integration kernel $F_2$, and using the definition of linear power
spectrum, we obtain,
\begin{eqnarray}
 \langle \delta^3 (z) \rangle
  & = & 6 D^4(z) \int \frac{d\veck_1}{(2\pi)^3}
  \int \frac{d\veck_2}{(2\pi)^3}
  F_2^{(s)}(\veck_1, \veck_2)
  \nonumber\\&&{}\times
  P_{\rm lin}(\veck_1) P_{\rm lin}(\veck_2).
\end{eqnarray}
After integrating over the directions of wavevectors, $\hatk_1$ and
$\hatk_2$, we find
\begin{equation}
 \langle \delta^3(z) \rangle = 
  D^4(z) \left(\frac{34}{7} + \frac{8 g_\ast^2}{175}\right)
  \left[\int \frac{dk k^2}{2\pi^2} A_{\rm lin}(k)\right]^2.
\end{equation}
On the other hand, the variance $\langle \delta^2 \rangle$
can be computed in a similar manner, and to the same order, we obtain
\begin{equation}
 \langle \delta^2(z) \rangle = 
  D^2(z) \int \frac{d\veck}{(2\pi)^3}P_{\rm lin}(\veck)
  = D^2(z) \int \frac{dk k^2}{2\pi^2}A_{\rm lin}(k).
\end{equation}
Therefore, we find the skewness to be
\begin{equation}
  S_3 = \frac{\langle \delta^3(z) \rangle}{\langle \delta^2(z)
  \rangle^2} = \frac{34}{7} + \frac{8g_\ast^2}{175}.
\end{equation}
The requirement that the power spectrum be positive definite
restricts the value $0 < g_\ast < 1$, and so the largest possible
deviation from the isotropic value of $34/7$ is less than
$1\%$.  Thus, the skewness will not be an effective
discriminant for anisotropic power.

To conclude, we used third-order perturbation theory to calculate the
quasilinear evolution of a primordial power spectrum.  Our
results show that nonlinear evolution suppresses the primordial
anisotropy.
The suppression is $\sim$7\% at $k = 0.1$~Mpc$^{-1}$ and $z = 0$
compared with the linear theory.
This must be taken into account when interpreting
the result of searches in the quasilinear regime for primordial
anisotropy, and it indicates that nonlinear scales are still
valuable as tests for the isotropy of primordial power, since the
suppression is weak.  We also found that the standard prediction for
the skewness is little changed if the primordial power spectrum has
a quadrupole anisotropy.

Our perturbative results seem to indicate that a quadrupole anisotropy
is enhanced at smaller scales, but this enhancement cannot be trusted
at low redshifts, as it occurs at scales where our perturbative
approach breaks down.  It seems counterintuitive to think that growth
of perturbations in the highly nonlinear regime can act to enhance the
primordial anisotropy, but it will require an N-body simulation to be
sure.  Likewise, we imagine that departures from statistical isotropy
that are higher order in angle (octupole, etc.) will also be
suppressed by quasilinear evolution, but we leave that calculation for
future work.

Before closing, we provide explicit expressions for the
quantities in Eq.~(\ref{eq:completepowerspectrum}).
For the numerical evaluation of Eqs.~(\ref{eq:P_22}) and
(\ref{eq:P_13}), we shall choose $\veck$ along the $z$-direction,
$\hatk = (0,0,1)$, and the preferred direction $\bm{\hat e}$ in the
$x$-$z$ plane, $\bm{\hat e} = (\sqrt{1-\mu_k^2}, 0, \mu_k)$.
We then use the spherical coordinate for $\vecq = q (\sqrt{ 1 - \mu^2
} \cos \phi, \sqrt{ 1 - \mu^2 } \sin \phi, \mu)$, and $d\vecq = q^2
dq d\mu d\phi$.
With this notation, $B_{mn}$ in Eq.~(\ref{eq:completepowerspectrum}) are
given as
\begin{eqnarray}
  B_{mn} & = & S_{mn} + 2 T_{mn},
  \\
  S_{mn} &=& \frac{k^4}{\pi^2} \int_0^\infty dq \int_{-1}^1 d\mu
  \frac{[7k\mu+q(3-10\mu^2)]^2}{(k^2+q^2-2kq\mu)^3}
  \mathcal I_{mn}
  \nonumber\\&&{}\times
  A_{\rm lin}(q)
  A_{\rm lin}\left(\sqrt{k^2+q^2-2kq\mu}\right),
  \\
  T_{mn} &=& \frac{k^2A_{\rm lin}(k)}{\pi^2}
  \int_0^\infty dq A_{\rm lin}(q)
  \biggl[\mathcal J_{mn}
  \nonumber\\&&{}
    + \mathcal K_{mn}
    (q^2-k^2)^3 \ln\left(\frac{k+q}{|k-q|}\right)\biggr],
\end{eqnarray}
where $S_{mn}$ and $T_{mn}$ correspond to $P_{22}$ and $P_{13}$,
respectively.
Explicit forms of $\mathcal I_{mn}$, $\mathcal J_{mn}$, and
$\mathcal K_{mn}$ are
\begin{eqnarray}
  \mathcal I_{00} &=& \frac{k^2+q^2-2kq\mu}{392},
  \\
  \mathcal I_{12} &=&
  \frac{2q^2(3\mu^2-1)+(k^2-2kq\mu)(3\mu^2+1)}{784},
  \\
  \mathcal I_{20} &=& \frac{2q^2-4kq\mu+k^2(3\mu^2-1)}{3920},
  \\
  \mathcal I_{22} &=&
  \frac{(k^2+q^2)(3\mu^2-1)-kq\mu(3\mu^2+1)}{2744},
  \\
  \mathcal I_{24} &=&
  \frac{9}{54880}\left[4k^2(3\mu^2-1)+q^2(35\mu^4-30\mu^2+3)
    \right.\nonumber\\&&{}\left.
  -8kq\mu(5\mu^2-3)\right],
  \\
  \mathcal J_{00} &=& \frac{1}{1008}
  \left(6\frac{k^2}{q^2}-79+50\frac{q^2}{k^2}-21\frac{q^4}{k^4}\right),
  \\
  \mathcal J_{12} &=& \frac{1}{40320}
  \left(90\frac{k^4}{q^4}+375\frac{k^2}{q^2}-5006+1732\frac{q^2}{k^2}
    \right.\nonumber\\&&{}\left.
  -300\frac{q^4}{k^4}-315\frac{q^6}{k^6}\right),
  \\
  \mathcal J_{20} &=& \frac{7}{10}\mathcal J_{22} = \frac{7}{18}
  \mathcal J_{24}
  \nonumber\\&=&
  \frac{1}{201600}\left(90\frac{k^4}{q^4}+135\frac{k^2}{q^2}-1846
  -268\frac{q^2}{k^2}
  \right.\nonumber\\&&{}\left.
  +540\frac{q^4}{k^4}-315\frac{q^6}{k^6}\right),
  \\
  \mathcal K_{00} &=& \frac{7q^2+2k^2}{672k^5q^3},
  \\
  \mathcal K_{12} &=& \frac{6k^6+41k^4q^2+76k^2q^4+21q^6}{5376k^7q^5},
  \\
  \mathcal K_{20} &=& \frac{7}{10}\mathcal K_{22} = \frac{7}{18}
  \mathcal K_{24}
  \nonumber\\&=&
  \frac{6k^6+25k^4q^2+20k^2q^4+21q^6}{26880k^7q^5},
\end{eqnarray}
and all the other components vanish.
Note that $T_{00}$ is the same as Eq.~(19) of Ref.~\cite{Jain:1994}.

\acknowledgments

MK acknowledges the hospitality of the Aspen Center for Physics,
where part of this work was completed.  This work was supported
by the Sherman Fairchild Foundation (SA), DoE DE-FG03-92-ER40701, NASA
NNG05GF69G, and the Gordon and Betty Moore Foundation (MK).


\begin{thebibliography}{}

\bibitem{preferreddirections} 
  C.~Gordon, W.~Hu, D.~Huterer and T.~Crawford,
  Phys.\ Rev.\  D {\bf 72}, 103002 (2005);
  J.~G.~Cresswell, A.~R.~Liddle, P.~Mukherjee and A.~Riazuelo,
  Phys.\ Rev.\  D {\bf 73}, 041302 (2006);
  S.~H.~S.~Alexander,
  hep-th/0601034;
  A.~Berera, R.~V.~Buniy and T.~W.~Kephart,
  JCAP {\bf 0410}, 016 (2004);
  R.~V.~Buniy, A.~Berera and T.~W.~Kephart,
  Phys.\ Rev.\  D {\bf 73}, 063529 (2006);
  G.~V.~Chibisov and Yu.~V.~Shtanov,
  Int.\ J.\ Mod.\ Phys.\  A {\bf 5}, 2625 (1990);
  G.~V.~Chibisov and Yu.~V.~Shtanov,
  Sov.\ Phys.\ JETP {\bf 69}, 17 (1989)
  [Zh.\ Eksp.\ Teor.\ Fiz.\  {\bf 96}, 32 (1989)];
  A.~E.~Gumrukcuoglu, C.~R.~Contaldi and M.~Peloso,
  JCAP {\bf 0711}, 005 (2007);
  C.~Armendariz-Picon,
  JCAP {\bf 0709}, 014 (2007);
  J.~F.~Donoghue, K.~Dutta and A.~Ross,
  astro-ph/0703455;
  R.~A.~Battye and A.~Moss,
  Phys.\ Rev.\  D {\bf 74}, 041301 (2006);
  C.~G.~Bohmer and D.~F.~Mota,
  arXiv:0710.2003 [astro-ph];
  T.~S.~Pereira, C.~Pitrou and J.~P.~Uzan,
  JCAP {\bf 0709}, 006 (2007).

\bibitem{Ackerman:2007nb}
  L.~Ackerman, S.~M.~Carroll and M.~B.~Wise,
  Phys.\ Rev.\  D {\bf 75}, 083502 (2007).

\bibitem{Pullen:2007}
  A.~R.~Pullen and M.~Kamionkowski,
  Phys.\ Rev.\  D {\bf 76}, 103529 (2007).

\bibitem{sdss} {\tt http://www.sdss.org}

\bibitem{2dF} {\tt http://www.aao.gov.au/2df}

\bibitem{Goroff:1986}
  M.~H.~Goroff, B.~Grinstein, S.~J.~Rey and M.~B.~Wise,
  Astrophys.\ J.\  {\bf 311}, 6 (1986).

\bibitem{Makino:1992}
  N.~Makino, M.~Sasaki and Y.~Suto,
  Phys.\ Rev.\  D {\bf 46}, 585 (1992).

\bibitem{Jain:1994}
  B.~Jain and E.~Bertschinger,
  Astrophys.\ J.\  {\bf 431}, 495 (1994).

\bibitem{Kamionkowski:1998fv}
  M.~Kamionkowski and A.~Buchalter,
  Astrophys.\ J.\  {\bf 514}, 7 (1999).

\bibitem{Eisenstein:1999}
  D.~J.~Eisenstein and W.~Hu,
  Astrophys.\ J.\  {\bf 511}, 5 (1999).

\bibitem{Spergel:2006}
  D.~N.~Spergel {\it et al.}, 
  Astrophys.\ J.\ Suppl.\  {\bf 170}, 377 (2007).

\bibitem{Jeong:2006xd}
  D.~Jeong and E.~Komatsu,
  Astrophys.\ J.\  {\bf 651}, 619 (2006).

\bibitem{pee80} P.~J.~E.~Peebles, {\it The
     Large-Scale Structure of the Universe} (Princeton
     Univ. Press, Princeton, 1980).

\bibitem{Fry:1983cj}
  J.~N.~Fry,
  Astrophys.\ J.\  {\bf 279}, 499 (1984).


\end{thebibliography}
\end{document}